\documentclass{article}

\usepackage{microtype}
\usepackage[ruled,vlined]{algorithm2e}
\usepackage{graphicx}
\usepackage{subfigure}
\usepackage{booktabs} 
\usepackage{amsmath} 

\usepackage{hyperref} 

\usepackage[accepted]{icml2019}
\usepackage{lineno}
\usepackage{amssymb}
\usepackage{amsmath}
\usepackage{amsthm}
\usepackage{epsfig}
\usepackage{graphicx}
\usepackage{graphics}
\usepackage{float}
\usepackage{multirow}
\usepackage{bbm}
\usepackage{color}
\usepackage{lineno}
\usepackage[normalem]{ulem} 
\usepackage{makeidx}
\usepackage{xspace}

\newtheorem{theorem}{Theorem}

\newtheorem{Definition}{Definition}
\newtheorem{corollary}{Corollary}

\newtheorem{lemma}{Lemma}
\newtheorem{claim}{Claim}

\definecolor{darkred}{rgb}{1, 0.1, 0.3}
\definecolor{darkblue}{rgb}{0.1, 0.1, 1}
\definecolor{darkgreen}{rgb}{0,0.6,0.5}

\newcommand {\mm}[1] {\ifmmode{#1}\else{\mbox{\(#1\)}}\fi}

\newcommand{\norm}[1]{\left\lVert#1\right\rVert}

\begin{document}

\twocolumn[
\title{N-grams Bayesian Differential Privacy}
\author{Osman Ramadan, Microsoft, osramada@microsoft.com \\ James Withers, Microsoft, jawithe@microsoft.com \\ Douglas Orr\footnotemark, Graphcore, douglas.orr99@gmail.com}
\date{\vspace{-0.2in}}
\maketitle

\begin{abstract}
Differential privacy has gained popularity in machine learning as a strong privacy guarantee, in contrast to privacy mitigation techniques such as k-anonymity. However, applying differential privacy to n-gram counts significantly degrades the utility of derived language models due to their large vocabularies. We propose a differential privacy mechanism that uses public data as a prior in a Bayesian setup to provide tighter bounds on the privacy loss metric $\epsilon$, and thus better privacy-utility trade-offs. It first transforms the counts to log space, approximating the distribution of the public and private data as Gaussian. The posterior distribution is then evaluated and softmax is applied to produce a probability distribution. This technique achieves up to 85\% reduction in KL-divergence compared to previously known  mechanisms at $\epsilon$ equals 0.1.  We compare our mechanism to k-anonymity in a n-gram language modelling task and show that it offers competitive performance at large vocabulary sizes, while also providing superior privacy protection.
\\

\textbf{Keywords:} Differential Privacy, n-grams, k-anonymity, NLP.
\end{abstract}
\vspace{0.4in}
]

\footnotetext[1]{During his time at Microsoft}
\section{Introduction}\label{sec:introduction}
In today's data-driven world, protecting the privacy of individuals' information is of the utmost importance to data curators, both as an ethical consideration and as a legal requirement, e.g. Article 29 of the European Union's General Data Protection Regulation describes privacy risks as singling out, linkability and inference.

Sequential data, such as DNA sequences, textual data and mobility traces, is being increasingly used in a variety of real-life applications, spanning from genome and language modeling to location-based recommendation systems. However, using such data poses considerable threats to individual privacy. It might be used by a malicious adversary to discover potential sensitive information about a data owner such as their habits, religion or relationships. 

Data anonymisation is a popular means of privacy preservation in datasets. One such example is the K-anonymity framework \cite{knon}, \cite{dp-any}, which anonymises data by generalising quasi identifiers, ensuring that an individual's data is indistinguishable from at least ($k-1$) others'. However, even the K-anonymity approach still poses privacy concerns, since it is deterministic and susceptible to privacy attacks, such as linkage attacks. It is therefore urgent to respond to the failure of existing anonymisation techniques by developing new schemes with provable privacy guarantees. 

Differential privacy is one of the only schemes that can be used to provide such guarantees. The main idea of differential privacy is to add noise to operations performed on a dataset so that an adversary cannot decide whether a particular user is included in the dataset or not. Due to the inherent sequentiality and high-dimensionality of sequential data, it is challenging to apply differential privacy. In particular, naively adding noise to the occurrence counts of each distinct sequence or n-gram in the dataset negatively impacts the utility \cite{seq-dp}. Other approaches rely on applying differential privacy to the operation of set union \cite{union-set} by collecting a subset of items from each user, taking the union of such subsets, and disclosing the items whose noisy counts rise above a certain threshold. However, these mechanisms do not learn the counts or the probability distribution of these items, which requires adding significantly large noise. One possible direction for mitigating the curse of dimensionality is to leverage Bayesian learning, utilising public data to shape our prior of the private data \cite{bay-dp-sample}. Learning a Bayesian model typically involves sampling from a posterior distribution, therefore the learning process is inherently randomized.

In this paper, we model applying differential privacy as taking a sample from the posterior distribution in a Bayesian setup. We propose a practical mechanism that efficiently uses public data to improve the utility and reduce the privacy loss. The paper is organised as follows. Section \ref{sec:preliminaries} gives a gentle introduction to differential privacy, defining core concepts and mechanisms that achieve differential privacy. Section \ref{sec:bayes-approx} provides a walk-through our proposed mechanism, and Section \ref{sec:results} discusses the experimental results of our mechanism in comparison with other mechanisms. Finally, Section \ref{sec:bounds} tries to empirically evaluate the privacy loss of our mechanism by performing membership inference attacks. Proofs of the theorems, claims and corollaries are included in the Appendices. 

\section{Preliminaries}\label{sec:preliminaries}
In this section, we review the definition of $(\epsilon, \delta)$-differential privacy, and the exponential mechanisms to achieve it.
\subsection{Differential Privacy}\label{subsec:dp}
Differential privacy is a statistical guarantee of privacy protection. It renders individuals' information indistinguishable by adding noise \cite{dp}, \cite{dp2}.
Let $\mathcal{V}$ be the set of the attributes and $\mathcal{U}$ the set of users their information we want to protect, then $\mathcal{D} = (\mathbf{c}^1, ..., \mathbf{c}^{|\mathcal{U}|})$ is a database of counts where each row $\mathbf{\Delta c}^n \in \mathbb{R}^{|\mathcal{V}|}$ corresponds to a user $n \in \mathcal{U}$ and $\Delta c_{i}^{n}$ is user $n$ count of attribute $i \in \mathcal{V}$, i.e. $\mathbf{c} = \sum_{n \in \mathcal{U}}\mathbf{\Delta c}^{n}$. We refer to the attributes set $\mathcal{V}$, which is independent of the users, as the vocabulary and the attributes as words, but they can equally refer to n-grams. The user adjacency is defined such that $\mathcal{D}$ and $\mathcal{D}_{-n}$ are two adjacent datasets, in which $\mathcal{D}_{-n}$ does not include user $n$.
\begin{Definition}\label{def:dp} (Differential Privacy)
Let $\epsilon > 0$ and $\delta \geq 0$ be given privacy parameters. We say a randomized mechanism $\mathcal{M}: \mathcal{D} \rightarrow \mathcal{M}(\mathcal{D})$ satisfies $(\epsilon, \delta)$-differential privacy, if for any adjacent datasets $\mathcal{D}$ and $\mathcal{D}_{-n}, \; n \in \mathcal{U}$ and all measurable sets $\mathcal{S} \subseteq Range(\mathcal{M})$, we have:
$$
Pr[\mathcal{M}(\mathcal{D}) \in \mathcal{S})] \leq e^{\epsilon} Pr[\mathcal{M}(\mathcal{D}_{-n})\in \mathcal{S}] + \delta 
$$
\end{Definition}
To apply differential privacy to a function $f$ we need to know its sensitivity:
\begin{Definition}\label{def:sen}($l_j$-sensitivity)
The $l_j$-sensitivity of a function $f: \mathcal{D} \rightarrow f(\mathcal{D}) \in \mathbb{R}^d$ is:
$$ \norm{\Delta f}_j = \max_{n \in \mathcal{U}}\norm{f(\mathcal{D}) - f(\mathcal{D}_{-n})}_j
$$

\end{Definition}
\subsection{The Exponential Mechanism}\label{subsec:expo}
We define two standard exponential mechanisms that achieve differential privacy, specifically the Laplace and Gaussian mechanisms \cite{dp}, \cite{dp2}.
\begin{Definition}\label{def:lap} (Laplace Mechanism)
For any function $f: \mathcal{D} \rightarrow \mathbb{R}^d$, the mechanism:
$$\mathcal{M}_{lap}(\mathcal{D}, f, \epsilon) = f(\mathcal{D}) + \mathbf{e}$$
$$e_i \sim Lap(\frac{\epsilon}{\norm{\Delta f}_1})$$
gives $(\epsilon, 0)$-differential privacy, where $Lap(b)$ is a b-scale Laplace distribution and $\norm{\Delta f}_1$ is the $l_1$-sensitivity of $f$.
\end{Definition}
\begin{Definition}\label{def:gau} (Gaussian Mechanism)
For any function $f: \mathcal{D} \rightarrow \mathbb{R}^d$, the mechanism:
$$\mathcal{M}_{Gau}(\mathcal{D}, f, \epsilon, \delta) = f(\mathcal{D}) + \boldsymbol{\eta}$$
$$\eta_i \sim \mathcal{N}(0, \sigma^2)$$
gives $(\epsilon, \delta)$-differential privacy, for $\epsilon \in (0, 1)$ and $\sigma > \sqrt{2ln(1.25/\delta)}\frac{\norm{\Delta f}_2}{\epsilon}$.
\end{Definition}
\section{Bayesian Approach to Utilise Public Data}\label{sec:bayes-approx}
Let $\boldsymbol\alpha$ be the public counts, defined as a one row database, and the private database $\mathcal{D}$ counts are defined over the public database vocabulary $\mathcal{V}$. To combine public and private counts in Bayesian differential privacy settings, we model the public data as the prior and the private data as the likelihood. Then, we compute the posterior and take a sample from it as the output  \cite{privacy-free}.
In the simple case, the public counts can be modelled as a Dirichlet distribution and the private counts as a Multinomial distribution \cite{bay-ngram}, \cite{bay-dp-bsc}. However, to form a differential private mechanism we must truncate the Dirichlet distribution at the tails to prevent the sensitivity of the private counts from exploding \cite{bay-dp-sample}, \cite{bay-dp-bsc}. Instead, we can approximate the Dirichlet distribution of the public counts in the softmax space as a Gaussian distribution $\mathcal{N}(\mathbf{h}|\boldsymbol{\mu}_p, \sigma_p\mathbf{I};\boldsymbol{\alpha})$ \cite{lap-approx}, \cite{vae} where $\sigma_p$ is the standard deviation and:
\begin{equation}\label{equ:public}
    \mu_{pi} = \log(\alpha_i + 1) - \frac{1}{|\mathcal{V}|}\sum_{j}^{|\mathcal{V}|}\log(\alpha_j + 1)
\end{equation}
which represents the prior distribution and the addition of 1 is important to avoid running into infinity. Using the private counts as a likelihood distribution in the Bayesian settings requires applying the softmax, i.e. the posterior distribution is $\sigma(\mathbf{h})\mathcal{N}(\mathbf{h}|\boldsymbol{\mu}_p, \sigma_p\mathbf{I})$. However, this is not a closed form posterior and so a differentially-private iterative sampling algorithm is required \cite{dp-wo-sen}. To mitigate this problem, we use a trick by modelling the overall private counts as one sample in the softmax/log space as follows, where $c_i$ is the total counts of word $i$:
\begin{equation}\label{equ:private}
    \hat{x}_i = \log(c_i + 1) - \frac{1}{|\mathcal{V}|}\sum_{j}^{|\mathcal{V}|}\log(c_j + 1)
\end{equation}
The adjacent database with user $n$ removed is defined as:
\begin{equation}\label{equ:private-adj}
    (\hat{x}_{-n})_{i} = \log(c_i - \Delta c^{n}_{i} + 1) - \frac{1}{|\mathcal{V}|}\sum_{j}^{|\mathcal{V}|}\log(c_j - \Delta c^{n}_{j} + 1) 
\end{equation}
Consequently, the likelihood can be represented by a Gaussian distribution: $\mathcal{N}(\mathbf{x}|\mathbf{h}, \sigma_l\mathbf{I})$, where $\sigma_l$ does not depend on the private database.
Since a Gaussian is a conjugate prior to itself, the posterior is also a Gaussian distribution: $\mathcal{N}(\mathbf{h}|\boldsymbol{\mu}_{ps}, \sigma_{ps}\mathbf{I}; \boldsymbol{\mu}_p, \boldsymbol{\hat{x}}, \sigma_p, \sigma_l)$, where $\boldsymbol{\mu}_{ps} = \rho \boldsymbol{\hat{x}} + (1 - \rho) \boldsymbol{\mu}_{p}$, $\sigma_{ps} = \rho \sigma_p$ and $\rho = \frac{\sigma_l}{\sigma_l + \sigma_p}$ is a hyperparameter to be chosen. We then apply the softmax to get the normalised probability distribution:
\begin{equation}\label{equ:softmax}
    \theta_i = softmax(\mathbf{h})_i = \frac{e^{h_i}}{\sum_{j}^{|\mathcal{V}|}e^{h_j}}
\end{equation}

\begin{theorem}\label{thm:bayes}
Given the Gaussian mechanism $\mathcal{M}$:= $\mathcal{N}(\mathbf{h}|\rho \boldsymbol{\hat{x}} + (1 - \rho) \boldsymbol{\mu}_{p}, \sigma_{ps}\mathbf{I})$ and its $l_2$-sensitivity: \\ $\norm{\Delta\boldsymbol{\hat{x}}}_{2} = \norm{\boldsymbol{\hat{x}} - \boldsymbol{\hat{x}}_{-n}}_{2} \leq \frac{\epsilon \sigma_{ps}}{\rho\sqrt{2\log(\frac{1.25}{\delta})}} \: \forall n \in \mathcal{U}$, then mechanism $\mathcal{M}$ is ($\epsilon, \delta$)-differentially private, where $\boldsymbol{\mu}_{p}$, $\boldsymbol{\hat{x}}$ and $\boldsymbol{\hat{x}}_{-n}$ are defined by equations \ref{equ:public}, \ref{equ:private} and \ref{equ:private-adj} respectively.
\end{theorem}
\subsection{Word-wise Adaptive Clipping Strategy \cite{adaclip}}\label{subsec:clip}
We examine the $l_2$-sensitivity $\Delta\boldsymbol{\hat{x}}$ in Theorem \ref{thm:bayes}:
$$
\norm{\Delta\boldsymbol{\hat{x}}}_{2}^{2} = \sum_{i}^{|\mathcal{V}|}\bigg(\log(1 - \frac{\Delta c^{n}_{i}}{1  + c_i}) - \frac{1}{|\mathcal{V}|}\sum_{i}^{|\mathcal{V}|}\log(1 - \frac{\Delta c^{n}_{j}}{1  + c_j}) \bigg)^{2}
$$
We observe from the equation above that the sensitivity depends mostly on words with large user percentage contribution ($\frac{\Delta c^{n}_{i}}{1  + c_i}$). That means frequent words have little effect on the sensitivity as opposed to rare words, even though they contribute the most to the utility, e.g. KL divergence. To mitigate the effect of rare words, we define a decay function $w(N_i) \in [0, 1]$ where $N_i$ is the number of users that have the word $i$ in their counts. A good decay function will penalise rare words more as fewer users contributed these words and so their sensitivity effect is large. The probability mass of these rare words can be taken from the public distribution. We experimented with clamped logarithmic, exponential and linear decay functions, and they all behaved similarly, therefore we decided to pick the linear function as it is the simplest and has a constant gradient:
\begin{equation}\label{equ:weight}
    \mathbf{w}_i = min\big(1, \frac{SN_i}{C}\big), \forall i \in \mathcal{V}
\end{equation}
Where $C$ is the maximum per-user word count and $S$ is a weighting hyperparameter.
\begin{corollary}\label{cor:weight-bayes}
Let $\boldsymbol{\mu}_{ps} = \rho\mathbf{w}\odot\boldsymbol{\hat{x}} + (1 - \rho) \boldsymbol{\mu}_{p}$ then the \\ modified Gaussian mechanism $\mathcal{M}$:= $\mathcal{N}(\mathbf{h}|\boldsymbol{\mu}_{ps}, \sigma_{ps}\mathbf{I})$ is ($\epsilon, \delta$)-differentially private if $\norm{\Delta\boldsymbol{\mu}_{ps}}_2 = \norm{\mathbf{w}\odot\mathbf{\Delta\hat{x}} + \frac{S}{C}\mathbbm{1}\Big(\mathbf{N} \leq \frac{C}{S}\Big)\odot\mathbf{\hat{x}}_{-n}}_{2} \leq \frac{\epsilon \sigma_{ps}}{\rho\sqrt{2\log(\frac{1.25}{\delta})}}\\
\forall n \in \mathcal{U}$ where $\mathbbm{1}$ is the indicator function, $\odot$ is element-wise multiplication and $\mathbf{w}, S$ and $C$ are defined in equation \ref{equ:weight}. 
\end{corollary}
The $l_2$-sensitivity can be computed by a brute-force method that is $~O(W.|U|)$, where $W$ is the maximum number of words each user can contribute and $|\mathcal{U}|$ is the number of users in the private database. The $l_2$-sensitivity can also be bounded provided $C$, $S$ and the private counts $\mathbf{\hat{x}}$. 
\begin{claim}\label{clm:bound}
Given $S$, $C$, $\mathbf{\hat{x}}$ and the number of users for each word $\mathbf{N}$, then the sensitivity defined in corollary \ref{cor:weight-bayes} is bounded as follows:
$$
    \norm{\Delta\boldsymbol{\mu}_{ps}}_2 < \norm{\mathbf{w}\odot\mathbf{L}C + \frac{S}{C} |\mathbf{\hat{x}}| \odot \mathbbm{1}\Big(\mathbf{N} \leq \frac{C}{S}\Big)}_{2} \forall n \in \mathcal{U}
$$
$$
\mathbf{L}_i = (1-\frac{1}{|\mathcal{V}|})\frac{1}{N_i} + \sum_{j\neq i}^{|\mathcal{V}|} \frac{1}{|\mathcal{V}|N_j}
$$

\end{claim}

\begin{algorithm}[H]\label{alg:bayes}\small
\SetKwInOut{Input}{Input}
\SetAlgoLined
\Input{$\{\mathbf{\Delta c}^{n}\}_{1}^{|\mathcal{U}|}$, $\boldsymbol{\alpha}$, $\epsilon$, $\delta$, $S$, $C$, $\rho$, $|\mathcal{V}|$}
\KwResult{normalised counts $\boldsymbol{\theta}$}
 \For{$n\leftarrow 1$ \KwTo $|\mathcal{U}|$}{
    $\mathbf{\Delta c}^{n} \gets \min(C, \mathbf{\Delta c}^{n})$; \tcp{clamp counts to C}
 }
 $\mathbf{c} \gets \sum_{n \in \mathcal{U}}\mathbf{\Delta c}^{n}$; \tcp{overall word counts}
 $\mathbf{N} \gets$ \texttt{no\_users\_per\_word}$(\{\mathbf{\Delta c}^{n}\}_{1}^{|\mathcal{U}|})$;
 
 $\mathbf{\hat{x}} \gets \log(\mathbf{c} + 1) - \frac{1}{|\mathcal{V}|}\norm{\log(\mathbf{c} + 1)}_1$; \tcp{equation \ref{equ:private}}
 $\mathbf{w} \gets \min(1, \frac{S\mathbf{N}}{C})$; \tcp{equation \ref{equ:weight}}
 $\mathbf{r} \gets \mathbf{w} \odot \mathbf{\hat{x}}$\;
 
 \eIf{Using the brute-force method to estimate sensitivity}{
 $\gamma \gets 0$; \tcp{initial sensitivity}
  \For{$n\leftarrow 1$ \KwTo $|\mathcal{U}|$}{
    $\gamma \gets \max\Big[\gamma, \norm{\mathbf{w}\odot\Delta\boldsymbol{\hat{x}} + \frac{S}{C}\mathbbm{1}(\mathbf{\Delta c}^{n}>0)\boldsymbol{\hat{x}}_{-n}}_{2}\Big]$\;
  }
 }{
  Compute worst-case estimate of the sensitivity ($\gamma$); \tcp{claim \ref{clm:bound}}
 }
 $\sigma_{ps} \gets \frac{\rho\gamma\sqrt{2\log(\frac{1.25}{\delta})}}{\epsilon}$; \tcp{corollary \ref{cor:weight-bayes}}
 $\boldsymbol{\mu}_p \gets \log(\boldsymbol{\alpha} + 1) - \frac{1}{|\mathcal{V}|}\norm{\log(\boldsymbol{\alpha} + 1)}_1$; \tcp{equation \ref{equ:public}}
 $\boldsymbol{\mu}_{ps} \gets \rho \boldsymbol{\hat{x}} + (1 - \rho) \boldsymbol{\mu}_{p}$\;
 
 Sample $\mathbf{h} \sim \mathcal{N}(\mathbf{h}; \boldsymbol{\mu}_{ps}, \sigma_{ps}\mathbf{I})$\;
 
 $\boldsymbol{\theta} \gets softmax(\mathbf{h})$\;
 \caption{BayesianDP}
\end{algorithm}
\subsection{Differentially Private Hyperparameter Tuning}
\label{subsec:private-tune}
$S$, $C$ and $\rho$ are hyperparameters and their values directly affect the ratio between the public and private counts and the variance of the posterior distribution for a given privacy parameters ($\epsilon, \delta$). If we ignore the noise we can compare the mean output of $\mathcal{M}$ after applying the softmax, i.e $softmax(\rho\mathbf{w}\odot\boldsymbol{\hat{x}} + (1 - \rho) \boldsymbol{\mu}_{p})$, to the normalised private counts distribution $softmax(\boldsymbol{\hat{x}})$ using KL-divergence. In other words, KL-divergence, for given private and public counts, is a function of $S$, $C$ and $\rho$. Consequently, there are optimal values $S = S^{*}, C= C^{*}, \rho = \rho^{*}$ at which the KL-divergence is minimised for any given public and private databases. $\rho$ is the weighting between private and public counts and $S$ and $C$ affect the amount of probability mass redistributed from frequent to rare words. A very small $S$ or very large $C$ transform the private counts to a uniform distribution. However, early experiments showed that $C$ has little effect on the KL-divergence unless the number of users is very large and/or there are rare words with high per-user counts, the latter is a clear privacy violation, so we do not vary $C$ and fix it to either 1 or 10 in all our experiments. Consequently, we only have two hyperparameters $S, \rho$ to vary and a grid-search hyperparameter tuning approach can be used at an extra cost of a small privacy budget (since we are only adding noise to scalar value) using a variant of the noisy max algorithm \cite{private-selection}, \cite{private-selection2}.

The idea of private hyperparameter tuning is to split the private counts into two non-overlapping sets, evaluate the KL-divergence between the normalised counts ($\frac{\mathbf{c}}{\norm{\mathbf{c}}_1}$) of the second set and the mean ($\boldsymbol{\mu}_{ps}$) of the Gaussian mechanism applied to the first set for different hyperparameters values ($S, \rho$). Then, add Laplace noise with privacy loss $\epsilon_2$ to each KL-divergence score and find the hyperparameter values ($S^{*}, \rho^{*}$) that give the minimum noisy score. Finally, using the mean ($\boldsymbol{\mu}_{ps}^{*}$) at these values, we sample from the ($\epsilon, \delta$)-DP Gaussian mechanism and report the final privacy parameters ($\epsilon + \epsilon_2, \delta$) using the strong composition theorem \cite{dp}. However, estimating the sensitivity of KL-divergence can be very complex and it is easier to use the un-normalised cross-entropy since they both exhibit their global minimum at the same hyperparameter values:

\begin{equation}\label{equ:cross-entropy}
    q(\mathbf{c},  \boldsymbol{\mu}_{ps}) = -\sum_{i}^{|\mathcal{V}|}c_i\big(\mu_{psi} - \log\sum_{j}^{|\mathcal{V}|}\exp{\mu_{psj}}\big)
\end{equation}
\begin{theorem}\label{thm:hyper-tuning}
Given two non-overlapping sets of private counts, let $\mathbf{c}$ be the counts of the first set and $\{\boldsymbol{\mu}_{ps}\}_{k=1}^{K}$ be $K$ evaluations of the mean of the Gaussian mechanism in Corollary \ref{cor:weight-bayes}, applied to the second set, for different hyperparameter values $(S_k, \rho_k)$. Then for any $l$-Lipschitz continuous scoring function: $q_k = q(\mathbf{c},  \boldsymbol{\mu}_{ps}^{(k)}) \: \forall k \in K$, the mechanism: $k^{*} = argmin_{k \in K} \big(q_k + e \sim Lap(\frac{\epsilon}{2|\Delta q|})\big)$, is $\epsilon$-differential private, where $Lap(b)$ is a $b$-scale Laplace distribution and $|\Delta q|$ is the maximum of the scoring function sensitivities with respect to each set separately, i.e $|\Delta q| =$ \scalebox{0.97}{$\max\Big[\big|q(\mathbf{c}, \boldsymbol{\mu}_{ps}) - q(\mathbf{c}_{-n}, \boldsymbol{\mu}_{ps})\big|, \big|q(\mathbf{c}, \boldsymbol{\mu}_{ps}) - q(\mathbf{c}, \boldsymbol{\mu_{ps}}_{-n})\big|\Big]$}
\end{theorem}

As mentioned before, using un-normalised cross-entropy as a scoring function instead of KL-divergence simplifies the sensitivity analysis while achieving the same result. 
\newpage
\begin{claim}\label{clm:cross-entropy}
If the scoring function in Theorem \ref{thm:hyper-tuning} is defined by Equation \ref{equ:cross-entropy}, then sensitivity $|\Delta q|$ is bounded as follows: \\
$ |\Delta q| \leq$ \\ \scalebox{0.8}{$\max\bigg[\rho\norm{\mathbf{c}}_2 \norm{\Delta\boldsymbol{\mu}_{ps}}_2, \max_{n \in \mathcal{U}_1}\bigg(\sum_{i}^{|\mathcal{V}|}\Delta c_{i}^{n}\Big(\mu_{psi} - \log\sum_{j}^{|\mathcal{V}|}\exp{\mu_{psj}}\Big)\bigg)\bigg]$} \scalebox{0.9}{$< \max\bigg[\rho\norm{\mathbf{c}}_2 \norm{\Delta\boldsymbol{\mu}_{ps}}_2, C_1\sum_{i}^{|\mathcal{V}|}\Big|\mu_{psi} - \log\sum_{j}^{|\mathcal{V}|}\exp{\mu_{psj}}\Big|\bigg]
$}
where $\mathcal{U}_1$ is the first private set and $C_1$ is their maximum word count. 
\end{claim}
The bounds in Claim \ref{clm:cross-entropy} allow us to reuse the estimated sensitivity of the second private set $\norm{\Delta\boldsymbol{\mu}_{ps}}_2$ when sampling from ($\epsilon, \delta$)-differentially private Gaussian mechanism.

\begin{algorithm}[H]\small\label{alg:hyper}
\SetKwInOut{Input}{Input}
\SetAlgoLined
\Input{$\{\mathbf{\Delta c}_{1}^{n}\}_{1}^{|\mathcal{U}_1|}$, $\{\mathbf{\Delta c}_{2}^{n}\}_{1}^{|\mathcal{U}_2|}$, $\boldsymbol{\alpha}$, $\epsilon_1$, $\epsilon_2$, $\delta$, $\{S_k, \rho_k\}_{1}^{K}$, $C$, $|\mathcal{V}|$}
\KwResult{normalised counts $\boldsymbol{\theta}, \epsilon_1 + \epsilon_2, \delta$}

 Clamp the counts and compute $\mathbf{c_1}, \mathbf{N_2}, \boldsymbol{\mu}_p, \mathbf{\hat{x}_2}$ \tcp{Extracted from algorithm 1}\;
 \For{$k\leftarrow 1$ \KwTo $K$}{
     $\boldsymbol{\mu}_{psk}, \sigma_{psk} \gets$ \texttt{BayesianDP(}$\mathbf{\hat{x}_2}, \boldsymbol{\mu}_p, \mathbf{N_2}, S_k, \rho_k, C, \epsilon_2, \delta$\texttt{)}\;
     
     $q_k \gets q(\mathbf{c_1}, \boldsymbol{\mu}_{psk}) + e \sim Lap(\frac{\epsilon_1}{2|\Delta q|})$; \tcp{$q$ from equation \ref{equ:cross-entropy} and $|\Delta q|$ from claim \ref{clm:cross-entropy}}
 }
 $i \gets argmin_{k \in K}(q_k)$\;
 
 Sample $\mathbf{h} \sim \mathcal{N}(\mathbf{h}; \boldsymbol{\mu}_{psi}, \sigma_{psi}\mathbf{I})$\;
 
 $\boldsymbol{\theta} \gets softmax(\mathbf{h})$\;
 \caption{EndToEndDP}
\end{algorithm}
\section{Experiments and Results}\label{sec:results}
\subsection{Dataset}
For all experiments we used the Reddit data, which consists of 4.4M users, as the private dataset and the Google Billion Word \cite{billion-word} as the public dataset. We extracted \textit{trigrams} and their counts from these datasets to form our database. Accordingly, the vocabulary $\mathcal{V}$ is the total number of distinct trigrams, which is determined only by the public dataset, and the output distribution of our mechanism is the joint trigram distribution or $P(w_1, w_2, w_3)$.
\subsection{Baselines}
We used two simple baselines in our experiments:
\begin{enumerate}
    \item The standard Laplace mechanism \cite{dp}, \cite{dp2}, where we allow each user to contribute up to $W$ counts overall, and add Laplace noise with scale $b = W/\epsilon$, then threshold at 0 and normalise to get a valid probability distribution.
    \item The other baseline is a modified Laplace mechanism that utilises public counts. We first normalise the public count, then for each user independently, we normalise the user counts and subtract the normalised public counts from it. After that, we get the weighted average of the subtracted output over all the users and add Laplace noise to it. Finally we add the normalised public counts, threshold at 0 and re-normalise. In other words, the un-normalised output of the modified Laplace mechanism:
$$\max\Big[\sum_{n \in \mathcal{U}}{\big[w_{n}(\mathbf{\Delta\Tilde{c}}^{n} - \boldsymbol{\Tilde{\alpha}})\big]} + e + \boldsymbol{\Tilde{\alpha}}, \mathbf{0}\Big]
$$
Where $\boldsymbol{\Tilde{\alpha}}$ is the normalised public counts, $\mathbf{\Delta\Tilde{c}}^{n}$ and $w_{n}=\frac{\Delta \mathbf{c}^n}{\mathbf{c}}$ are the normalised counts and weighting of user $n$ and $e \sim Lap(\frac{\epsilon}{\norm{\Delta f}_1})$ is the Laplace noise and $\norm{\Delta f}_1$ is the $l_1$-sensitivity of the modified mechanism.
\end{enumerate}

\subsection{Experimentation setup}\label{subsec:exp-setup}
For all experiments, the allowed word counts per user $C$ is set to 1 if the number of users is less than 1M, otherwise $C=10$. We also limit the total number of counts for each user to $|\mathcal{U}|/1000$. These values were chosen empirically from early experiments to be good utility-privacy trade-offs. For our Bayesian mechanism, we used the brute-force method to compute the sensitivity as it gives tighter bounds. In hyperparameter tuning, we split the private data into $90\%$ train set and $10\%$ validation set, where we spend $\frac{2\epsilon}{3}$ on training and $\frac{\epsilon}{3}$ on private hyperparameter tuning, and $S \in [10^{-3}, 10^{4}]$.
\subsection{Performance under different settings}
\begin{figure}[h]
\begin{center}
\includegraphics[width=0.95\linewidth]{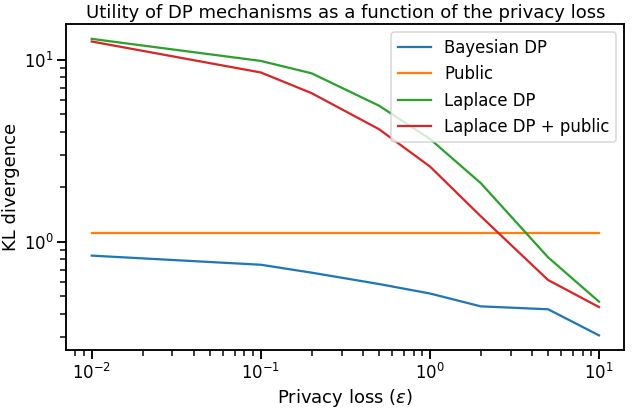}
\caption{KL-divergence from the private distribution vs $\epsilon$ for different mechanisms. The number of users is 4M and vocabulary size is 50K.}
\label{fig:kl-eps}
\end{center}
\end{figure}

\begin{figure*}[ht]
\begin{center}
\centerline{\includegraphics[width=0.92\textwidth]{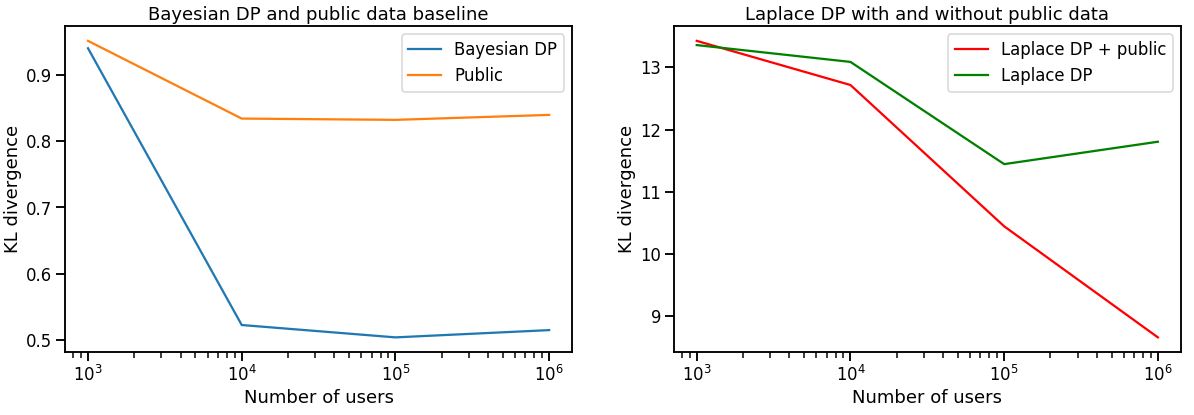}}
\caption{Left: KL-divergence from the private distribution vs the number of users for Bayesian differential privacy compared to a public baseline. Right: Laplace differential privacy with and without public data. $\epsilon = 0.1$ and vocabulary size is 50K.}
\label{fig:kl-users}
\end{center}
\end{figure*}

\begin{figure*}[h!]
\begin{center}
\includegraphics[width=0.92\textwidth]{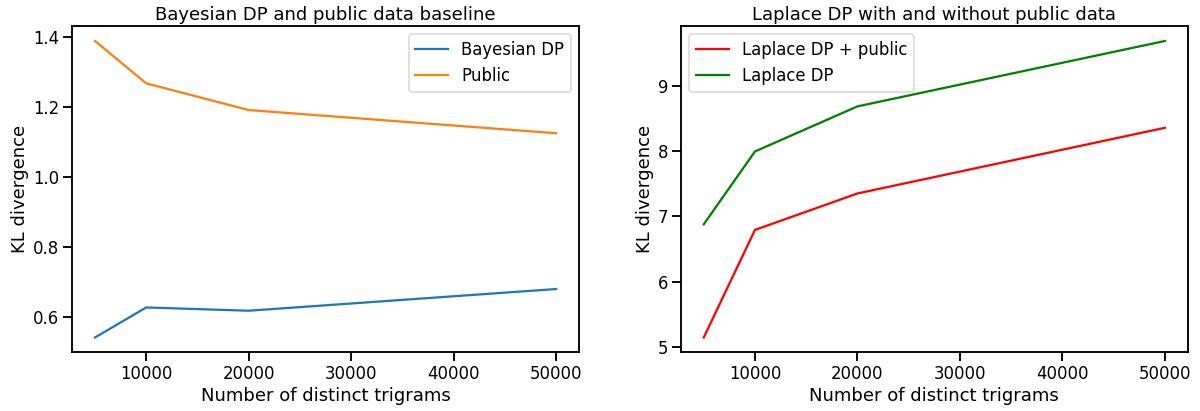}
\caption{Left: KL-divergence from the private distribution vs the number of distinct trigrams (vocabulary size) for Bayesian differential privacy compared to a public baseline. Right: Laplace differential privacy with and without public data. The number of users is 4M and $\epsilon = 0.1$.}
\label{fig:kl-vocab}
\end{center}
\end{figure*}

\begin{figure}[h!]
\begin{center}
\includegraphics[width=0.95\linewidth]{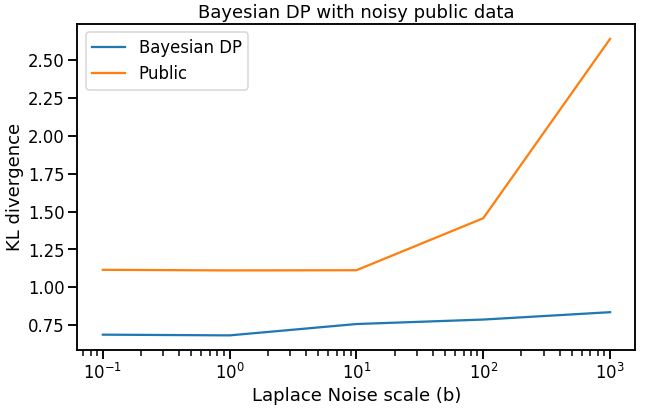}
\caption{KL-divergence of Bayesian differential privacy vs the scale of the Laplace noise added to the public data to deteriorate its quality. The number of users is 4M, $\epsilon = 0.1$ and $|\mathcal{V}| = 50K$.}
\label{fig:kl-public}
\end{center}
\end{figure}

We evaluate the utility of our mechanism, measured as the KL-divergence from the private data distribution, under different settings such as the number of users in private database, the size of vocabulary set and the quality of the public data. We also compare its utility with other baseline mechanisms, as shown in Figures \ref{fig:kl-eps} to \ref{fig:kl-public}.

\subsection{Comparison with K-Anonymity}\label{subsec:k-anon}
\begin{figure*}[h!]
\begin{center}
\includegraphics[width=0.95\textwidth]{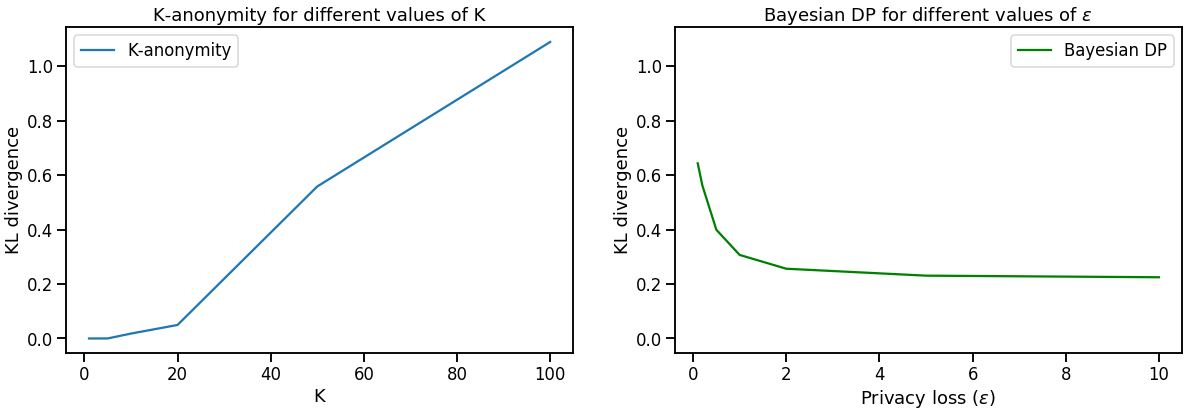}
\caption{Left: KL-divergence from the private distribution of K-anonymity vs K. Right: KL-divergence of Bayesian DP vs $\epsilon$. Both plots have the same scale on the y-axis (KL-divergence). The number of users is 100K and vocabulary size is 50K.}
\label{fig:kl-kan}
\end{center}
\end{figure*}

We compare our mechanism to K-anonymity, by evaluating the KL-divergence of the method at different values of K as shown in Figure \ref{fig:kl-kan}. For $\epsilon=0.2$, we achieve similar utility to K-anonymity with $K=50$. However, differential privacy offers superior privacy guarantees.

\subsection{Performance in Language Modeling}\label{subsec:lm}
We compare the perplexity of our mechanism applied to n-grams in language modeling to the other baselines and K-anonymity. Since the performance of n-gram language models depends on the back-off strategies and  how out-of-vocabulary words are handled, we constrain the data to only in-vocabulary words and sentences covered by trigrams so no back-off strategy is required, leading to a fair comparison in as shown in Table \ref{tab:per}. 

\begin{table}[!h]

\vskip 0.05in
\begin{center}
\begin{small}
\begin{sc}
\begin{tabular}{lcccr}
\toprule
 \textbf{Mechanism} & \textbf{Perplexity} \\
\midrule
Private baseline&29.48\\ 
K-anonymity with K = 50&35.51\\ 
Bayesian DP at $\epsilon=0.2$&35.55\\
Laplace DP + public data at $\epsilon=0.2$&580.46\\ 
Laplace DP at $\epsilon=0.2$&894.32\\ 
Public baseline&43.00\\
\bottomrule
\end{tabular}
\end{sc}
\end{small}
\end{center}
\caption{Average word perplexity for different mechanisms using trigram language models. The number of distinct n-grams is 60K}\label{tab:per}
\vskip 0.1in
\end{table}

\section{Empirical Privacy Bounds By Membership Inference Attack}\label{sec:bounds}
\begin{figure}[H]
\begin{center}
\includegraphics[width=0.95\linewidth]{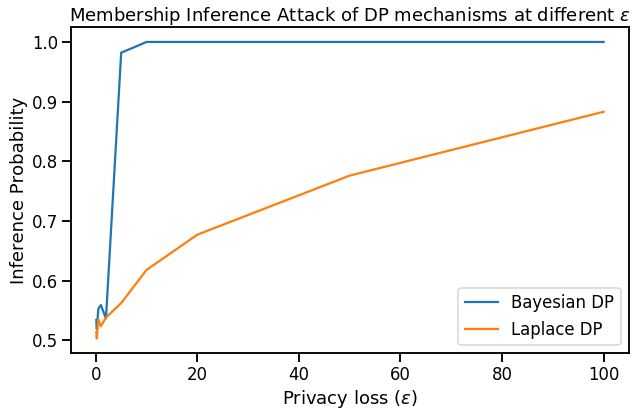}
\caption{Membership inference probability vs $\epsilon$ for Laplace DP and Bayesian DP. The number of users is 1.2M and the vocabulary size is 50K.}
\label{fig:mem-eps}
\end{center}
\end{figure}

 The reason why our Bayesian mechanism is outperforming the standard Laplace mechanism for the same $\epsilon$ is because our mechanism has tighter bounds on the privacy loss. This is due to the word-based clipping strategy that minimizes the sensitivity effect of rare words. Moreover, the brute-force method of estimating the sensitivity tied with the hyperparameter tuning allows us to find the optimal privacy vs utility trade-off. To confirm this, we apply a membership inference attack \cite{eval-dp} to the output of our Bayesian mechanism and compare against the Laplace mechanism. To perform the attack, we apply the mechanism to two adjacent datasets, one of which does not have the most contributing, in terms of user counts. We then sample two probability distributions $\boldsymbol{\theta}, \boldsymbol{\theta_{-u}}$ from each dataset and repeat the same process $N$ times. The probability of membership inference is then the number of times the KL-divergence, relative to the removed user distribution, of $\boldsymbol{\theta}$ is less than that of $\boldsymbol{\theta_{-u}}$ divided by $N$. Plots in Figure \ref{fig:mem-eps} show that the probability of membership inference of the Laplace mechanism grows slowly with $\epsilon$, even at $\epsilon=100$ the probability is still less than $0.9$, whereas the inference probability of the Bayesian mechanism approaches $1$ as soon as $\epsilon \geq 5$. Therefore, the Bayesian mechanism gives a more accurate estimate of the true privacy loss $\epsilon$, since from the definition of differential privacy, a $\epsilon \geq 5$ indicates that the ratio between the output probability of the adjacent datasets is $> \exp(5) \approx 148$, which is a clear privacy violation. In other words, the Laplace mechanism is excessively pessimistic about $\epsilon$ and it is possible to achieve much higher utility at a lower $\epsilon$ using the Bayesian mechanism.
\section{Conclusion}
\label{conclusion}
In this paper, we proposed a novel Bayesian differential privacy approach, applied to n-grams, which utilises public data to provide a significantly better utility vs privacy trade-off compared to well-known privacy mechanisms, such as the Laplace mechanism. In our approach, we transform the counts to log space, approximating the distribution of the public and private data as Gaussian. The posterior distribution is then evaluated and softmax is applied to produce a probability distribution. We performed several experiments on n-grams from the Reddit dataset to demonstrate the superior performance of our mechanism, achieving up to 85\% reduction in KL-divergence compared to the Laplace mechanism for privacy loss $\epsilon = 0.1$, and similar KL-divergence performance to K-anonymity with $K=50$. Finally, we applied a membership inference attack to explain the improvement of our mechanism over the Laplace mechanism. The attack showed that our mechanism provides tighter bounds on the privacy loss and thus gives better estimate of $\epsilon$. Future work will investigate using a similar Bayesian approach to apply differential privacy to deep neural networks during Stochastic Gradient Descent (SGD) training.

\section*{Acknowledgements}
The authors of this paper would like to acknowledge the SwiftKey Task and Intelligence Research team, in particular Joe Osborne and Dmitry Stratiychuk, for their insights and suggestions that guided the research. We would also like to acknowledge everyone that reviewed this paper and provided invaluable guidance and feedback.

\bibliographystyle{ieeetr}
\bibliography{ref}

\newpage
\onecolumn
\appendix

\section{Proof of Theorem \ref{thm:bayes} and Corollary \ref{cor:weight-bayes}}\label{appendix:thm:bayes}
We start by proving Theorem \ref{thm:bayes} where the Gaussian mechanism proposed can be composed into a deterministic function $f$ and an additive Gaussian noise $\boldsymbol{\eta}$.\\
Define $f: \mathcal{N}^{|\mathcal{U}| \times V} \rightarrow \mathbb{R}^{V}$, where $\mathcal{U}$ is the set of all private users of a database $D$ and $V = |\mathcal{V}|$ is the vocabulary size. Adopting the same notation used in theorem \ref{thm:bayes}, let $D_{-n}$ be an adjacent database to $D$ with user $n \in |\mathcal{U}|$ removed and let $\boldsymbol{\alpha} \in \mathbb{R}^{V}, \mathbf{\Delta c^{(j)}} \in \mathbb{R}^{V}$ be the public and user $j$ counts respectively. The adjacency, $sup(\mathbf{c}, \mathbf{c}_{-n})$, is expressed formally as $\mathbf{c} = \sum_{j}^{|\mathcal{U}|}\mathbf{\Delta c}^{(j)}$ and $ \mathbf{c}_{-n} = \sum_{j\neq n}^{|\mathcal{U}|} \mathbf{\Delta c}^{(j)}$. Consequently, $\mathcal{M}: f(\mathbf{c}, \boldsymbol{\alpha}, \rho) + \boldsymbol{\eta}$, where:
\begin{align*}
    &f(\mathbf{c}, \boldsymbol{\alpha}, \rho) = \rho \mathbf{\hat{x}} + (1-\rho)\boldsymbol{\mu}_p \\
    &\mathbf{\hat{x}} = \log(\mathbf{c} + 1) - \frac{1}{V} \norm{\log(\mathbf{c} + 1)}_1 \\
    &\boldsymbol{\mu}_p = \log\boldsymbol{\alpha} - \frac{1}{V} \norm{\log\boldsymbol{\alpha}}_1 \\
    & \eta_i \sim \mathcal{N}(0, \sigma_{ps}), \forall i \in V 
\end{align*}
$\mathcal{M}$ satisfies a ($\epsilon, \delta$)-DP Gaussian mechanism if the standard deviation $\sigma_{ps} > \sqrt{2\log(\frac{1.25}{\delta})}\frac{\boldsymbol{\Delta}_2f}{\epsilon}$, where
\begin{align*}
    \boldsymbol{\Delta}_2f &= \max_{sup(\mathbf{c}, \mathbf{c}_{-n})} \norm{f(\mathbf{c}, \boldsymbol{\alpha}, \rho) - f(\mathbf{c}_{-n}, \boldsymbol{\alpha}, \rho)}_2 \\&
    = \max_{sup(\mathbf{c}, \mathbf{c}_{-n})} \norm{\rho \mathbf{\hat{x}} + (1-\rho)\boldsymbol{\mu}_p - \rho \mathbf{\hat{x}}_{-n} - (1-\rho)\boldsymbol{\mu}_p}_2 \\&
    = \rho \max_{sup(\mathbf{c}, \mathbf{c}_{-n})}\norm{\mathbf{\hat{x}} - \mathbf{\hat{x}}_{-n}}_2
\end{align*} 
And rearranging this will give:
\begin{equation*}
    \max_{sup(\mathbf{c}, \mathbf{c}_{-n})}\norm{\mathbf{\hat{x}} - \mathbf{\hat{x}}_{-n}}_2 < \frac{\epsilon\sigma_{ps}}{\rho \sqrt{2\log(\frac{1.25}{\delta})}}
\end{equation*}

Corollary \ref{cor:weight-bayes} follows from Theorem \ref{thm:bayes} if we replace $\mathbf{\hat{x}}$ with $\mathbf{w}\odot \mathbf{\hat{x}}$, where $\mathbf{w} = \min[1,  \mathbf{N}S/C]$ as defined by equation \ref{equ:weight} and $\mathbf{w}_{-n} = \min\Big[1, \big(\mathbf{N} - \mathbbm{1}(\mathbf{\Delta c}_{-n} > 0)\big)S/C\Big]$, since each user can only affect the number of users for any word $N_i$ by 1. As a result:
\begin{align*}
    \boldsymbol{\Delta}_2f/\rho &= \max_{sup(\mathbf{c}, \mathbf{c}_{-n})}\norm{\mathbf{w}\odot\mathbf{\hat{x}} - \mathbf{w}_{-n}\odot\mathbf{\hat{x}}_{-n}}_2\\
    & = \max_{sup(\mathbf{c}, \mathbf{c}_{-n})}\norm{\mathbf{w}\odot(\mathbf{\hat{x}} - \mathbf{\hat{x}}_{-n}) + (\mathbf{w} - \mathbf{w}_{-n})\odot\mathbf{\hat{x}}_{-n}}_2\\
    &= \max_{sup(\mathbf{c}, \mathbf{c}_{-n})}\norm{\mathbf{w}\odot\mathbf{\Delta\hat{x}} + \frac{S}{C}\mathbbm{1}\Big(\mathbf{N} \leq \frac{C}{S}\Big)\odot\mathbf{\hat{x}}_{-n}}_2
\end{align*}
The identity function in the last equality stems from clamping the decay function at $1$, i.e the gradient is zero for $\mathbf{N}S/C > 1$ or $\mathbf{N} > C/S$. We note also that the first equality can be rearranged differently, which will come handy in the proving claim \ref{clm:bound}, as follows:
\begin{align}
    \max_{sup(\mathbf{c}, \mathbf{c}_{-n})}\norm{\mathbf{w}_{-n}\odot\mathbf{\hat{x}} - \mathbf{w}_{-n}\odot\mathbf{\hat{x}}_{-n}}_2 &= \max_{sup(\mathbf{c}, \mathbf{c}_{-n})}\norm{\mathbf{w}_{-n}\odot(\mathbf{\hat{x}} - \mathbf{\hat{x}}_{-n}) + (\mathbf{w} - \mathbf{w}_{-n})\odot\mathbf{\hat{x}}}_2\nonumber\\
    & = \max_{sup(\mathbf{c}, \mathbf{c}_{-n})}\norm{\mathbf{w}_{-n}\odot\mathbf{\Delta\hat{x}} + \frac{S}{C}\mathbbm{1}\Big(\mathbf{N} \leq \frac{C}{S}\Big)\odot\mathbf{\hat{x}}}_2 \label{equ:sens}
\end{align}

\section{Proof of Claim \ref{clm:bound}}\label{appendix:clm:bound}
Claim \ref{clm:bound} provides a worst-case estimate of the sensitivity $\norm{\boldsymbol{\Delta\mu}_{ps}}_2$, that does not require recomputing sensitivity of $D_{-n}$ for each user $n$. Firstly, we state the following lemma:
\begin{lemma}\label{lem:bound}
The sensitivity $|\mathbf{\Delta \hat{x}}|$ of $\mathbf{\hat{x}}$, which is defined in equation \ref{equ:private}, is bounded from above as follows:
\begin{align*}
    &|\mathbf{\Delta\hat{x}}| \leq \mathbf{L}C\\
    & L_i = \big(1 - \frac{1}{V}\big)\frac{1}{N_i} + \sum_{j\neq i}^{V}\frac{1}{VN_j} 
\end{align*}
\end{lemma}
Then, using this lemma, we start the proof from Equation \ref{equ:sens}:
\begin{align*}
    \max_{sup(\mathbf{c}, \mathbf{c}_{-n})}\norm{\mathbf{w}_{-n}\odot\mathbf{\Delta\hat{x}} + \frac{S}{C}\mathbbm{1}\Big(\mathbf{N} \leq \frac{C}{S}\Big)\odot\mathbf{\hat{x}}}_2 &\leq \max_{sup(\mathbf{c}, \mathbf{c}_{-n})}\norm{\mathbf{w}_{-n}\odot|\mathbf{\Delta\hat{x}}| + \frac{S}{C}\mathbbm{1}\Big(\mathbf{N} \leq \frac{C}{S}\Big)\odot|\mathbf{\hat{x}}|}_2 \\
    & \leq \max_{sup(\mathbf{c}, \mathbf{c}_{-n})}\norm{\mathbf{w}\odot|\mathbf{\Delta\hat{x}}| + \frac{S}{C}\mathbbm{1}\Big(\mathbf{N} \leq \frac{C}{S}\Big)\odot|\mathbf{\hat{x}}|}_2 \\
    & \leq \norm{\mathbf{w}\odot\mathbf{L}C + \frac{S}{C}\mathbbm{1}\Big(\mathbf{N} \leq \frac{C}{S}\Big)\odot|\mathbf{\hat{x}}|}_2
\end{align*}
The last two inequalities follow from the monotonicity of $\mathbf{w}$ and Lemma \ref{lem:bound} respectively. \\
Now we prove Lemma \ref{lem:bound}, by defining a strictly increasing function $\mathbf{\hat{y}}$ similar to $\mathbf{\hat{x}}$:
\begin{equation*}
    \hat{y}_i = \log(c_i + 1)(1-\frac{1}{V}) + \sum_{j\neq i} \log(c_j + 1), i \in V 
\end{equation*}

The function $\hat{y}_i(\mathbf{c})$ has two important properties:
\begin{itemize}
    \item It is a concave function, since it's a strictly increasing linear combination of $\log$
    \item $|\Delta \hat{x}_i| \leq |\Delta\hat{y}_i| $
\end{itemize}
From the first property we can conclude that $\hat{y}_i$ is $L_i$-Lipschitz over $\mathbf{c}$ with respect to the $\norm{.}_{\infty}$ if $\norm{\nabla\hat{y}_i}_{1} \leq L_i$, since $l_1$ and $l_\infty$ are dual norms. Moreover:
\begin{align*}
    \frac{\partial \hat{y}_i}{\partial c_j} &= 
        \begin{cases} 
          (1-\frac{1}{V})\frac{1}{c_j + 1} & j = i \\
          \frac{1}{V(c_j + 1)} & j \neq i
       \end{cases}\\
       & \leq
       \begin{cases} 
          (1-\frac{1}{V})\frac{1}{N_j + 1} & j = i \\
          \frac{1}{V(N_j + 1)} & j \neq i
       \end{cases}
\end{align*}
The inequality follows because the smallest word count is equal to the number of users that have the word. Consequently, the maximum $l_1$ norm of the gradient: $L_i = \big(1 - \frac{1}{V}\big)\frac{1}{N_i} + \sum_{j\neq i}^{V}\frac{1}{VN_j}$.
From the Lipschitz property:
\begin{equation*}
    |\Delta \hat{x}_i| \leq |\Delta\hat{y}_i| = |\hat{y}_i(\mathbf{c}) - \hat{y}_i(\mathbf{c}_{-n})| \leq L_i \norm{\mathbf{c} - \mathbf{c}_{-n}}_{\infty} \leq L_iC, i \in V 
\end{equation*}
\begin{equation*}
    |\mathbf{\Delta \hat{x}}| \leq \mathbf{L}C
\end{equation*}

\section{Proof of Theorem \ref{thm:hyper-tuning}}
\paragraph{
The first part of the theorem is a variant of the noisy max algorithm, with the relaxation that the score function does not have to be monotonic. To follow the same steps as the noisy max proof, we will prove the theorem for the maximum of the negative score function $q$. This is allowed because of the symmetry of the Laplace noise.} Let $s_k = -q(\mathbf{c}, \boldsymbol{\mu_{ps}^{(k)}})$ and $s_k' = -q(\mathbf{c}_{-n}, \boldsymbol{\mu_{ps}^{(k)}}_{-n})$ be the scores for the adjacent databases $D, D_{-n}$ for $ \forall n \in |\mathcal{U}|$ and $k \in K$. For any $k \in K$, fix $\boldsymbol{\eta}_{-k}$, a draw from $\big[Lap(2|\Delta q|/\epsilon)\big]^{K-1}$ used for all noisy scores except the k\textsuperscript{th} score. We will argue for each $\boldsymbol{\eta}_{-k}$ independently. \\
Define the notation $P(k|\zeta)$ to mean the probability that the output of the mechanism (index of the maximum noisy score) is $k$, conditioned on $\zeta$. \\
First, we show that $P(k|D, \boldsymbol{\eta}_{-k}) \leq e^{\epsilon}P(k|D_{-n}, \boldsymbol{\eta}_{-k}), \forall n \in |\mathcal{U}|$. Define
\begin{equation*}
    \eta^{*} = \min_{\eta_k} s_k + \eta_k > s_m + \eta_m \; \forall m \neq k
\end{equation*}

Since we want the maximum value, $k$ will be the output when the database is $D$ if and only if $\eta_k \geq \eta^{*}$. Let $\Delta s_i = s_i - s'_i \; \forall i \in K$, we have for all $1 \leq m \neq k \leq K$:
\begin{align*}
    s_k + \eta^{*} &> s_m + \eta_m \\
    s'_k + \Delta s_k + \eta^{*} &> s'_m + \Delta s_m + \eta_m \\
    s'_k + 2|\Delta q| + \eta^{*} \geq s'_k + &\Delta s_k - \Delta s_m + \eta^{*} > s'_m + \eta_m
\end{align*}
Thus, if $\eta_k \geq 2|\Delta q| + \eta^{*}$, then the k\textsuperscript{th} score will be the maximum when the database is $D_{-n}, \; \forall n \in |\mathcal{U}|$ and the noise vector is $(\eta_k, \boldsymbol{\eta}_{-k})$. The probabilities below are over the choice of $\eta_k \sim Lap(2|\Delta q|/\epsilon|)$:

\begin{align*}
    & P(\eta_k \geq 2|\Delta q| + \eta^{*}) \geq e^{-\epsilon} P(\eta_k \geq \eta^{*}) = e^{-\epsilon}P(k|D, \boldsymbol{\eta}_{-k})\\
    P(k|D_{-n}, \boldsymbol{\eta}_{-k}) \geq & P(\eta_k \geq 2|\Delta q| + \eta^{*}) \geq e^{-\epsilon} P(\eta_k \geq \eta^{*}) = e^{-\epsilon}P(k|D, \boldsymbol{\eta}_{-k})
\end{align*}
Rearranging and multiplying both sides by $e^{\epsilon}$:
\begin{equation*}
    P(k|D, \boldsymbol{\eta}_{-k}) \leq e^{\epsilon} P(k|D_{-n}, \boldsymbol{\eta}_{-k}) \; \forall n \in |\mathcal{U}|
\end{equation*}
Now we show $P(k|D_{-n}, \boldsymbol{\eta}_{-k}) \leq e^{\epsilon} P(k|D, \boldsymbol{\eta}_{-k}) \; \forall n \in |\mathcal{U}|$, following similar steps as above. Define 
\begin{equation*}
    \eta^{*} = \min_{\eta_k} s_k' + \eta_k > s_m' + \eta_m \; \forall m \neq k
\end{equation*}
$k$ will be the output(argmax noisy score) when the database is $D_{-n} \; \forall n \in |\mathcal{U}|$ if and only if $\eta_k \geq \eta^{*}$
we have for all $1 \leq m \neq k \leq K$:
\begin{align*}
    s'_k + \eta^{*} &> s'_m + \eta_m \\
    s'_k + \Delta s_k + \Delta s_m+ \eta^{*} &> s'_m + \Delta s_k + \Delta s_m + \eta_m \\
    s_k + 2|\Delta q| + \eta^{*} \geq s_k + &\Delta s_m - \Delta s_k + \eta^{*} > s_m + \eta_m
\end{align*}
Thus, if $\eta_k \geq 2|\Delta q| + \eta^{*}$, then the k\textsuperscript{th} score will be the maximum when the database is $D$ and the noise vector is $(\eta_k, \boldsymbol{\eta}_{-k})$. The probabilities below are over the choice of $\eta_k \sim Lap(2|\Delta q|/\epsilon|)$:

\begin{align*}
    & P(\eta_k \geq 2|\Delta q| + \eta^{*}) \geq e^{-\epsilon} P(\eta_k \geq \eta^{*}) = e^{-\epsilon}P(k|D_{-n}, \boldsymbol{\eta}_{-k})\\
    P(k|D, \boldsymbol{\eta}_{-k}) \geq & P(\eta_k \geq 2|\Delta q| + \eta^{*}) \geq e^{-\epsilon} P(\eta_k \geq \eta^{*}) = e^{-\epsilon}P(k|D_{-n}, \boldsymbol{\eta}_{-k})
\end{align*}
And so:
\begin{equation*}
    P(k|D_{-n}, \boldsymbol{\eta}_{-k}) \leq e^{\epsilon} P(k|D, \boldsymbol{\eta}_{-k}) \; \forall n \in |\mathcal{U}|
\end{equation*}
Finally, since:
\begin{equation*}
    k^{*} = argmin_{k \in K} \Big[q_k - e \sim Lap\big(\frac{\epsilon}{2|\Delta q|}\big)\Big] = argmax_{k \in K} \Big[-q_k + e \sim Lap\big(\frac{\epsilon}{2|\Delta q|}\big)\Big]
\end{equation*}
which is $\epsilon$-differentially private, it is trivial to show that $k^{*} = argmin_{k \in K} \Big[q_k + e \sim Lap\big(\frac{\epsilon}{2|\Delta q|}\big)\Big]$ is also $\epsilon$-differentially private, because the Laplace distribution is symmetric around $0$ and the above proof only requires the sensitivity of $q$ and no other assumption is made on the properties of $q$. 
\paragraph{The second part of the theorem is a direct consequence of having non-overlapping private sets, i.e any user removed in $D, D_{-n}$ can only be from one of the private sets and not the other, thus we can treat each set separately and take the maximum sensitivity of both.}

\section{Proof of Claim \ref{clm:cross-entropy}}
The sensitivity with respect to the first private set, $(\mathbf{c}, \mathbf{c}_{-n})$, is a brute-force evaluation of adding and removing each user in the set. This is bounded for all users if you take the maximum word count in the set $C_1$ as shown in the claim. Therefore, here we prove the upper bound of the sensitivity with respect to the second private set $(\boldsymbol{\mu_{ps}}, \boldsymbol{\mu_{ps}}_{-n})$. \\
Let $\mathbf{r} = \mathbf{w}\odot \mathbf{\hat{x}}$ so $\boldsymbol{\mu_{ps}} = \rho\mathbf{r} + (1 - \rho) \boldsymbol{\mu_{p}}$. We first note that score function $q(.,.)$ defined in equation \ref{equ:cross-entropy} is a convex function in $\mathbf{r}$ since it is a scaled version of KL-divergence. Consequently, $q(.,.)$ is $l$-Lipschitz over $\mathbf{r}$ with respect to the $l_2$ norm if $\norm{\nabla q}_2 \leq l$, since $l_2$ is self-dual. Moreover, for all $i \in V$:
\begin{align*}
    \Big|\frac{\partial q}{\partial r_i}\Big| &= \bigg|c_i\Big(\rho - \frac{\rho \exp\big(\rho r_i + (1-\rho)\mu_{pi}\big)}{\sum_{j=1}^{V}\exp\big(\rho r_j + (1-\rho)\mu_{pj}\big)} \Big)\bigg| \\
    &= \big|c_i\rho(1 - \theta_i)\big| \;\; \theta_i \in [0, 1] \\
    &\leq c_i \rho
\end{align*}
Thus $q(.,.)$ is $l$-Lipschitz with $l = \rho\norm{\mathbf{c}}_2$. The Lipschitz property implies that: \\$\Big|q(\mathbf{c}, \rho\mathbf{r} + (1 - \rho) \boldsymbol{\mu_{p}}) - q(\mathbf{c}, \rho\mathbf{r}_{-n} + (1 - \rho) \boldsymbol{\mu_{p}})\Big| \leq \rho\norm{c}_2\norm{\mathbf{\Delta r}}_2 = \rho\norm{c}_2\norm{\boldsymbol{\Delta\mu_{ps}}}_2$  $\forall n \in |\mathcal{U}_1|$


\end{document}